\documentclass[preprints,article,accept,pdftex,moreauthors]{Definitions/mdpi}
\firstpage{1}
\pubvolume{1}
\issuenum{1}
\articlenumber{0}
\pubyear{2026}
\copyrightyear{2026}
\datereceived{}
\daterevised{}
\dateaccepted{} 
\datepublished{ }

\usepackage{xspace}
\newcommand{\K}{$^{39}$K\xspace}
\newcommand{\Rb}{$^{87}$Rb\xspace}
\newcommand{\He}{$^{3}$He\xspace}

\newcommand{\Rr}{\mathcal{R}}
\newcommand{\Dphi}{\Delta\varphi}
\newcommand{\kR}{\kappa_{\theta}}
\newcommand{\kDphi}{\kappa_{\theta}^{\Dphi}}
\newcommand{\SNR}{\mathrm{SNR}}
\newcommand{\Nn}{\mathcal{N}}
\AddToHook{begindocument/before}{\ifdefined\pdfstringdefDisableCommands
    \pdfstringdefDisableCommands{\def\K{K-39}\def\Rb{Rb-87}\def\He{He-3}\def\Rr{R}\def\Dphi{Delta phi}\def\SNR{SNR}\def\Nn{N}\def\xi{ξ\textunderscore}\def\sim{∼}\def\gtrsim{≳}}\fi
}

\Title{Optimal Calibration-Free Observable for the Nucleon-Coupling Ratio in a Dual-Alkali Comagnetometer for Dark Matter~Searches}

\Author{Yossi Rosenzweig $^{\dagger}$, Yevgeny Kats $^{\dagger}$, Eli Sarid, Menachem Givon, Yonathan Japha, Ron Folman}

\address[1]{Department of Physics, Ben-Gurion University of the Negev, Beer-Sheva 8410501, Israel\\ yosefros@post.bgu.ac.il (Y.R.); katsye@bgu.ac.il (Y.K.); esarid@bgu.ac.il (E.S.); japhay@bgu.ac.il (Y.J.)}

\firstnote{These authors contributed equally to this work.}

\abstract{A dual-alkali single-cell \Rb--\K--\He\ comagnetometer can read an axion-like dark matter signal through two optical rotation channels, encoding the ratio $\Rr=\xi_n/\xi_p$ of the field's neutron and proton spin couplings in their relative response. The inter-species phase difference $\Dphi$ has been proposed as a calibration-free readout that is sensitive to $\Rr$. Treating the extraction of $\Rr$ as a statistical estimation problem, we show that the estimate of $\Rr$ depends on the two channels only through their complex ratio, which splits into $\Dphi$ and an amplitude ratio, of which only $\Dphi$ is insensitive to the relative gain and hence calibration free. For our choice of comagnetometer parameters, above $\sim\!100$~Hz the phase difference alone captures most of the coupling-ratio information. At lower frequencies $\Dphi$ is no longer near-sufficient. Below $\sim\!40$~Hz, the amplitude ratio would improve the precision on $\Rr$ by a factor of $\gtrsim2$. Recovering that information, however, requires the relative gain to be known sufficiently accurately, so $\Dphi$ remains the robust observable even where it is not the most informative one.}


\begin{document}

\section{Introduction}

The nature of dark matter remains one of the central open problems in physics: its gravitational presence is firmly established, yet its particle identity and its couplings to Standard Model fields are largely unconstrained~\cite{Cirelli:2024ssz}. One class of candidates is ultralight pseudoscalar (axion-like) particles~\cite{Kimball:2023vxk,Adams:2022pbo,ParticleDataGroup:2026aaa}. Because~such light particles need enormous occupation numbers to account for the local dark matter density, they behave as a classical field oscillating coherently over $\sim\!10^{6}$ periods of their Compton frequency~\cite{Graham:2013gfa,Lisanti:2021vij}. Such a field generically couples derivatively to the axial-vector currents of Standard Model fermions~\cite{Graham:2013gfa}; in the non-relativistic limit this acts on fermion spins like a pseudomagnetic field, but~with couplings that are generally not proportional to the magnetic moments. Alkali magnetometers and alkali--noble-gas comagnetometers are therefore leading probes of this and similar scenarios~\cite{Bloch:2019lcy,Afach:2021pfd,Lee:2022vvb,Bloch:2022kjm,Afach2023,Wei:2023rzs,Gavilan-Martin:2024nlo,Khamis:2024oqa,Wilson:2025lhq}.

If a candidate signal is observed, a~decisive follow-up question is how the exotic field couples to the different subatomic particles---protons, neutrons, and~electrons---since the pattern of couplings discriminates among microscopic models. Because~the electronic structure and the fractional proton and neutron spin content differ among elements and isotopes~\cite{Kimball2015}, a~field with given microscopic couplings acts on different atomic species through different effective strengths. Earlier work showed how to disentangle these couplings using a \emph{set} of distinct comagnetometers~\cite{Rosenzweig2024}. Ref.~\cite{Letter} proposed instead a single-cell scheme, reviewed below, in~which two alkalis interact with the same field. While an amplitude combination of the channels cancels the magnetic background, thus enabling the detection of weaker signals, the~phase difference can be used to extract the coupling ratio. The~present paper extends ref.~\cite{Letter} to the question of how well a candidate signal can be characterized, retaining the response model, baseline operating point, and~noise budget of that~work.

The remainder of the paper is organized as follows. Section~\ref{sec:review} reviews the dual-alkali scheme of ref.~\cite{Letter} and the response model used throughout. Section~\ref{sec:estimation} formulates the extraction of $\Rr$ as an estimation problem, with~the unknown signal amplitude and phase eliminated from the inference. Section~\ref{sec:facts} establishes three structural observations that fix how $\Rr$ is extracted from the two channels: a single channel is blind to $\Rr$; the coupling-ratio information is carried by the complex inter-channel ratio; and its phase and amplitude carry complementary parts of the information, of~which only the phase is calibration free. Section~\ref{sec:results} evaluates the attainable precision and the sufficiency of $\Dphi$ as a function of frequency and $\Rr$. Section~\ref{sec:systematics} derives the tolerances on the operating point and on instrumental systematics and shows how they can be met. Section~\ref{sec:conclusions} summarizes the results. Appendix~\ref{app:proof} proves the reduction of the two-channel Fisher information used in the~body.

\section{The Dual-Alkali Correlated~Measurement}
\label{sec:review}

Ref.~\cite{Letter} considers a single vapor cell containing two alkali species, \Rb\ and \K, together with \He, polarized jointly by hybrid spin-exchange optical pumping~\cite{Walker1997,Babcock2003}. The~two alkalis are probed simultaneously by two overlapping, linearly polarized beams, each detuned $\sim\!0.1$~THz to the blue of its own D1 line (770.1~nm for \K, 795.0~nm for \Rb). After~traversing the cell, each probe acquires a Faraday rotation angle proportional to the transverse spin polarization $P_x^{\,j}$ of the alkali it addresses~\cite{Seltzer2008}. The~two beams are separated by a beamsplitter, passed through a narrow-band wavelength filter in each polarimetry arm, and~demodulated by a shared lock-in amplifier, so that the channels share a common phase origin. The~three species evolve under coupled Bloch equations~\cite{Padniuk:2021dtr,Rosenzweig2024,Padniuk:2023uyv} in which each species' exotic coupling is
\begin{equation}
\xi_j=\eta_j\,(\sigma_n^j\,\xi_n+\sigma_p^j\,\xi_p)\,,
\label{eq:xij}
\end{equation}
where $\xi_{n,p}$ are the underlying couplings to neutron and proton spin, and~$\sigma_{n,p}^j$ describe the nuclear spin content of the atomic species $j$ provided in Table~\ref{tab:spin}. The~prefactor $\eta_j$ equals $q_j-1$ for the alkalis, where $q_j$ is the nuclear slowing-down factor, and~1 for the noble gas. We follow ref.~\cite{Letter} by considering an example in which the field couples to nucleon spins but not to electron spin,\endnote{Coupling to electrons need not be present since quarks and leptons are independent degrees of freedom.} so the single quantity that characterizes the internal structure of the coupling is the ratio
\begin{equation}
\Rr\equiv\frac{\xi_n}{\xi_p}\,.
\end{equation}
Table~\ref{tab:params} collects the parameters and noise budget of ref.~\cite{Letter}.

\begin{table}[b]
\centering
\caption{Fractional contributions of the neutron and proton spins $\sigma_n^j$ and $\sigma_p^j$ to~the total nuclear spin of the three species of the cell, in~the convention of ref.~\cite{Kimball2015}.}
\label{tab:spin}
\begin{tabularx}{\textwidth}{lCCL}
\toprule
\textbf{Species} & \boldmath{$\sigma_n$} & \boldmath{$\sigma_p$} & \textbf{Refs.} \\
\midrule
\K  & $0.034$ & $-0.131$ & \cite{Kimball2015,Engel:1995gw} \\
\Rb & $0.083$ & $0.251$  & \cite{Flambaum:2006ip} \\
\He & $0.87$  & $-0.027$ & \cite{Kimball2015,Friar:1990vx} \\
\bottomrule
\end{tabularx}
\end{table}
\unskip

\begin{table}[t]
\centering
\caption{Baseline operating parameters and noise budget of ref.~\cite{Letter}, adopted throughout this paper. The~noise entries are one-sided amplitude spectral densities referred to an equivalent magnetic~field.}
\label{tab:params}
\begin{tabularx}{\textwidth}{llc}
\toprule
Cell temperature & $T$ & $180\,^\circ\mathrm{C}$ \\
\He\ number density & $n_\mathrm{He}$ & $3$~amg \\
Alkali number densities & $n_\mathrm{Rb}$, $n_\mathrm{K}$ & $4.0$, $5.45\times10^{13}\,\mathrm{cm}^{-3}$ \\
Equilibrium longitudinal polarizations & $P_z^\mathrm{Rb}$, $P_z^\mathrm{K}$, $P_z^\mathrm{He}$ & $0.46$, $0.45$, $0.047$ \\
Alkali relaxation rates & $\Gamma_\mathrm{Rb}$, $\Gamma_\mathrm{K}$ & $600$, $600$~Hz \\
Optical pumping rate (\Rb\ only) & $R_p$ & $1200$~Hz \\
Compensation weighting & $\alpha$ & $0.578$ \\
\midrule
Correlated magnetic background, $f>20$~Hz & $\sqrt{S_B}$ & $6.2\,\mathrm{fT}/\sqrt{\mathrm{Hz}}$ \\
Differential readout floor at $100$~Hz & $\tilde\sigma$ & $0.22\,\mathrm{fT}/\sqrt{\mathrm{Hz}}$ \\
\bottomrule
\end{tabularx}
\end{table}

The central observation is that a magnetic field acts on both alkalis through the (essentially common) electronic gyromagnetic ratio $\gamma_e$, whereas an exotic field acts through the species-dependent couplings of Equation~\eqref{eq:xij}. Both fields also reach the alkalis indirectly, through the \He, whose driven magnetization couples back to \Rb\ and \K\ via their unequal Fermi contact factors; this shared route is therefore species dependent for either field. It is strong at low frequency and rolls off as the \He\ can no longer follow the drive. In~a standard comagnetometer~\cite{Allred2002,Kornack2002}, the effects of low-frequency transverse magnetic fields are canceled by operating at the self-compensation point, reached by applying a longitudinal magnetic field at $B_c$, while the effects of exotic fields survive. With~the dual-readout approach of ref.~\cite{Letter}, which relies only on the two alkalis sharing the same field, the~magnetic field effects can also be subtracted at high frequencies while leaving generic exotic signals~unsuppressed.

An observable of interest is the inter-species phase difference of the alkali transverse spin polarizations $P_x^{\,j}$ at the angular frequency $\omega=2\pi f$ of the candidate signal,
\begin{equation}
\Dphi(\omega;\Rr)\equiv\varphi_\mathrm{K}(\omega;\Rr)-\varphi_\mathrm{Rb}(\omega;\Rr)\,,
\qquad
\varphi_j\equiv\arg P_x^{\,j}(\omega;\Rr)\,.
\label{eq:dphi}
\end{equation}
For a magnetic drive, $\Dphi$ is at the milliradian level across the rejection band, which the differing alkali Larmor frequencies confine to tens of Hz and above. Above~$\sim\!50$~Hz the exotic response phase difference is one to three orders of magnitude above that baseline. Because~the two probe beams are detuned to the same side of their respective lines, the~proportionality factor between the measured optical rotation and $P_x^{\,j}$ is real and positive for both channels~\cite{Letter}, and~$\Dphi$ inherits a \emph{calibration-free}
 character: it is immune to probe intensity, photodetector responsivity, cell length, and~correlated density~fluctuations.

\begin{figure}[!t]
\centering
\includegraphics[width=\textwidth]{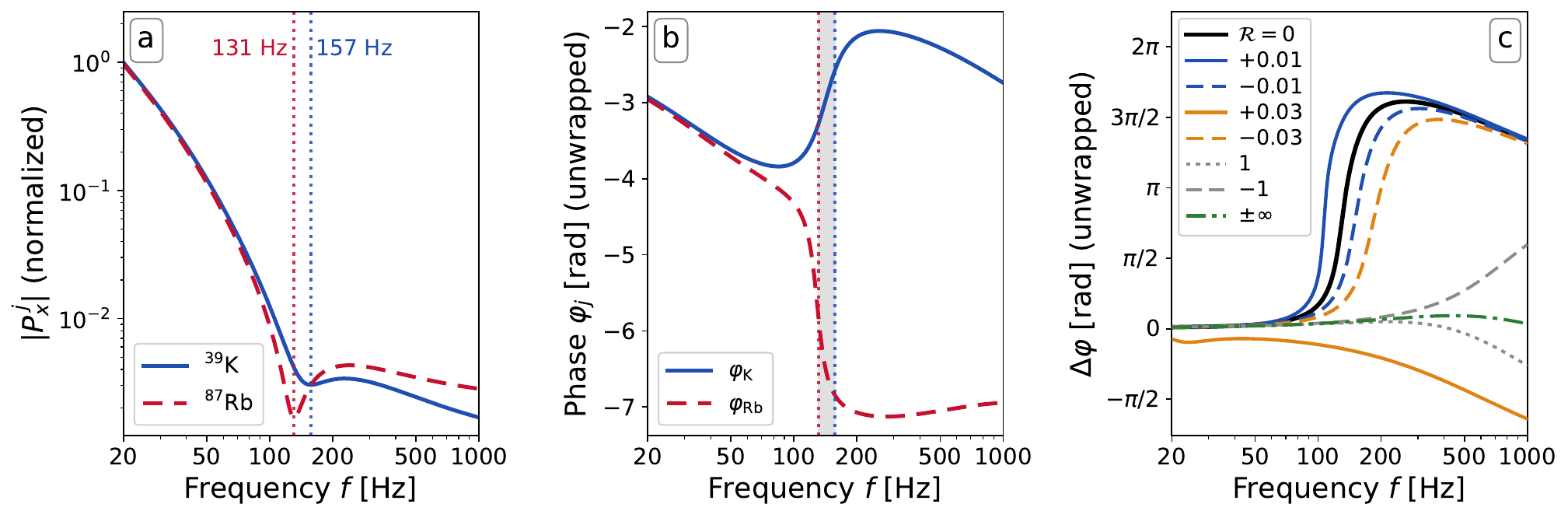}
\caption{Antiresonances in the exotic response of the two alkalis, and~the differential-phase swings they produce. (\textbf{a})~Magnitude of the exotic response $|P_x^{\,j}|$ for $j\in\{{}^{39}\mathrm{K},{}^{87}\mathrm{Rb}\}$ at a pure-proton coupling ($\Rr=0$), normalized to its largest value in the band: the destructive interference carves an antiresonance at $157$~Hz in \K and at $131$~Hz in \Rb (dotted vertical lines). (\textbf{b})~The corresponding unwrapped response phases $\varphi_j$. Across the interval between the two antiresonances (shaded) they wind in opposite senses, \K advancing by $+0.7$ and \Rb by $-1.0\,\mathrm{rad}$. (\textbf{c})~The resulting differential phase $\Dphi$ of Equation~\eqref{eq:dphi}, unwrapped, for~a family of small $|\Rr|$ of either sign, for~the equal mixtures $\Rr=\pm1$, and~for the pure-neutron limit $\Rr=\pm\infty$. The~swing is a single rapid step that both slides and shrinks with the coupling ratio. The~\K antiresonance moves from $235$~Hz at $\Rr=-0.03$ through $157$~Hz at $\Rr=0$ to $128$~Hz at $\Rr=+0.01$, and~disappears just above $\Rr=+0.03$, where the step is already gone. The~excursion is reduced from $5.0\,\mathrm{rad}$ at $\Rr=0$ to $1.8$, $1.0$ and $0.26\,\mathrm{rad}$ at $\Rr=-1$, $+1$ and $\pm\infty$, where no antiresonance is left~either.}
\label{fig:dphicuts}
\end{figure}

Two further features of ref.~\cite{Letter} are important to mention. First, the~longitudinal compensation field cannot self-compensate both alkalis at once, because~their Fermi contact factors differ; it is parameterized as a weighted compromise (Equation~(6) of ref.~\cite{Letter}) controlled by a single parameter $\alpha$, and~throughout this paper we adopt the baseline operating point $\alpha=0.578$ along with the other system parameters of ref.~\cite{Letter}. Second, at~a pure-proton coupling ($\Rr=0$) the destructive interference among the three exotic drive channels---direct, \He-mediated, and~alkali--alkali spin exchange---carves an \emph{antiresonance}
in the \K exotic response at $\sim\!157$~Hz and in the \Rb response at $\sim\!131$~Hz~\cite{Letter}. Both antiresonances persist, at~coupling-dependent frequencies, over~a range of $\Rr$ about this value. Near~these frequencies the two alkali phases wind in opposite senses, so $\Dphi$ accumulates a single rapid swing that encodes $\Rr$ (Figure~\ref{fig:dphicuts}). Distinct phase differences also arise where no antiresonance occurs. Mapping $\Dphi$ across the coupling-ratio--frequency plane acts as a matched filter for the coupling ratio. The~full map will be presented in Section~\ref{sec:results}.

\unskip

\section{The Estimation~Problem}
\label{sec:estimation}

Ref.~\cite{Letter} establishes $\Dphi$ as a practical, calibration-free readout that carries information about $\Rr$. It does not, however, address a logically prior question: among \emph{all} observables one could build from the two channels, which one extracts $\Rr$ with the smallest statistical variance, and~how much of that optimal information does the calibration-free $\Dphi$ actually contain? That is the question we answer~here.

Each measurement channel $j\in\{{}^{39}\mathrm{K},{}^{87}\mathrm{Rb}\}$ records a complex demodulated amplitude
\begin{equation}
z_j=\Nn\,\mathcal{K}_j\,T^b_j(\Rr)+n_j\,,
\label{eq:model}
\end{equation}
where $\mathcal{K}_j$ is the channel's optical-rotation gain (the real, positive prefactor of Equation~(1) of ref.~\cite{Letter}) and $T^b_j(\Rr)$ is the exotic transverse-spin transfer function.\endnote{The superscript $b$ refers to the exotic field, which is denoted by $\mathbf{b}$
 in ref.~\cite{Letter}; it is to be distinguished from the superscript $B$ of the magnetic transfer function $T^B_j$ introduced below.} The factor $\Nn$ is the external field, carrying unknown amplitude and phase, which drives both channels. It is therefore a common complex nuisance parameter. The~readout noise $n_j$ for each channel is circular: its two quadratures---the real and imaginary parts of the demodulated signal---are independent white Gaussian processes of equal one-sided amplitude spectral density $\sigma_j$, so that in a bandwidth $\Delta f$ each has root-mean-square $\sigma_j\sqrt{\Delta f}$.\endnote{Circularity follows from the demodulation, the~shot noise of a linear polarimeter having no preferred phase with respect to the lock-in reference. Independence between the channels follows from the separate probe beams and photodetectors, whose shot noise is uncorrelated. Gaussianity follows from the photon number per demodulated sample: a one-second sample carries more than $10^{12}$ photons even at a probe power of a microwatt. Whiteness is required only across the detection bandwidth $\Delta f$ about the signal frequency. The~probe modulation of ref.~\cite{Letter} places the detection band above the $1/f$ region of the polarimetry chain, where the spectrum is flat over any $\Delta f$ of interest.\label{foot:white}} The bandwidth $\Delta f$ is the reciprocal of the coherent integration time, and~no smaller than the linewidth of the field, $\sim\!10^{-6}f$ for a virialized dark matter field~\cite{Graham:2013gfa}.

The response is affine in the coupling ratio, i.e.,~it has the form $T^b_j(\Rr)=A_j+\Rr\,B_j$, since $\Rr$ enters only through the per-species drive $\xi_j(\Rr)\propto\sigma_n^j\Rr+\sigma_p^j$ of Equation~\eqref{eq:xij}. The~ratio $\Rr=\xi_n/\xi_p$ may lie anywhere on the real axis. As~it diverges for a pure-neutron coupling ($\xi_p\to0$), it is a poor coordinate near that limit. We will therefore use the singularity-free \emph{coupling angle} $\theta$, defined by $(\xi_p,\xi_n)=\xi\,(\cos\theta,\sin\theta)$, where the overall strength $\xi$ is absorbed into the amplitude $\Nn$ of Equation~\eqref{eq:model} and $\theta$ carries the coupling ratio alone, so that $\Rr=\tan\theta$. The~range $\theta\in[0,\pi)$ covers every physical coupling ratio exactly once: pure proton at $\theta=0$, equal couplings at $\theta=\pi/4$, pure neutron at $\theta=\pi/2$, and~opposite-sign couplings ($\Rr<0$) for $\theta\in(\pi/2,\pi)$. The~response is then
\begin{equation}
T^b_j(\theta)=A_j\cos\theta+B_j\sin\theta \,,
\label{eq:affine}
\end{equation}
where $A_j$ is the pure-proton and $B_j$ the pure-neutron response of channel $j$.

We will work throughout in terms of the standard estimation theory notions of Fisher information and the Cram\'er--Rao bound~\cite{Kay1993,LehmannCasella}. The~Fisher information $I_\lambda$ for a parameter $\lambda$ is the expected sharpness of the log-likelihood peak at the true parameter value, $I_\lambda=-\mathbb{E}\big[\partial_\lambda^{2}\ln L\big]$, equivalently the noise-ensemble variance of the first derivative $\partial_\lambda\ln L$: the more sharply peaked the log-likelihood, the~more the data constrain the parameter. Its reciprocal bounds the variance of any unbiased estimator---the Cram\'er--Rao bound, $\sigma_\lambda^{2}\geq1/I_\lambda$. Throughout we quote the resulting standard deviation, $\sigma_\lambda=1/\sqrt{I_\lambda}$; it is the attainable precision in the small-noise regime, and~the Cram\'er--Rao lower bound otherwise. The~Fisher information values below are written with the spectral densities in place of variances, i.e.,~for a $1$~Hz band. A band $\Delta f$ divides them by $\Delta f$, which cancels from every information ratio and every precision quoted at fixed amplitude signal-to-noise ratio (SNR).

\section{Structural Observations on the Extraction of $ \Rr$}
\label{sec:facts}

The following three observations about the dual-readout setup fix how $\Rr$ is extracted from the two channels and the standing of $\Dphi$ in that~extraction.

\subsection{A Single
 Channel Is Blind to $\Rr$}

With one complex measurement and $\Nn$ free, $\Rr$ and $\Nn$ are unidentifiable: any apparent change in $\Rr$ can be reabsorbed into a redefinition of $\Nn$. Extracting $\Rr$ fundamentally requires both~channels.

\subsection{The Coupling-Ratio Information Is Carried by the Complex Inter-Channel Ratio}

Since the common unknown factor $\Nn$ enters both channels alike, it drops out of the ratio $z_\mathrm{K}/z_\mathrm{Rb}$, leaving a quantity that is independent of $\Nn$ but still fully determined by $\Rr$. This ratio is complex; we work with its logarithm
\begin{equation}
\mathcal{Z}\equiv\ln\left(\frac{T^b_\mathrm{K}}{T^b_\mathrm{Rb}}\right)\,,
\end{equation}
which turns the multiplicative amplitude/phase structure into additive real and imaginary parts---$\mathrm{Re}\,\mathcal{Z}=\ln|T^b_\mathrm{K}/T^b_\mathrm{Rb}|$ the log-amplitude ratio and $\mathrm{Im}\,\mathcal{Z}=\varphi_\mathrm{K}-\varphi_\mathrm{Rb}=\Dphi$ the differential phase---so that the amplitude and phase observables sit on the two Cartesian axes and, as~shown below, split the information additively. With~independent readout noise in the two channels, each quadrature of the measured $\mathcal{Z}$ has variance $w/|\Nn|^{2}$, with~\begin{equation}
w=\left(\frac{\sigma_\mathrm{K}}{\mathcal{K}_\mathrm{K}\,|T^b_\mathrm{K}|}\right)^{\!2}+\left(\frac{\sigma_\mathrm{Rb}}{\mathcal{K}_\mathrm{Rb}\,|T^b_\mathrm{Rb}|}\right)^{\!2}\,.
\label{eq:weight}
\end{equation}
The unknown drive $\Nn$ is removed by profiling: the likelihood is maximized over $\Nn$ at each value of $\Rr$. Doing so on the full two-channel likelihood---exact for Gaussian, independent, circular per-channel noise---gives, after~an algebraic (Lagrange) reduction,\endnote{Profiling the Gaussian likelihood over $\Nn$ leaves the Fisher information for $\Rr$ as the projection of $\partial_\Rr\mu$ off the drive direction $\mu_j\equiv\mathcal{K}_jT^b_j$; for two channels Lagrange's identity collapses it to $|\partial_\Rr\mathcal{Z}|^2/w$; Appendix~\ref{app:proof} carries out the reduction in full, including the correlated-background generalization of Equation~\eqref{eq:weff}. Treating $\mathcal{Z}$ itself as Gaussian with variance $w$ (an approximation) happens to lead to the same expression as the exact approach.\label{foot:Lagrange}} the Fisher information for $\Rr$ per unit $|\Nn|^{2}$,\endnote{The complete information for $\Rr$ is $|\Nn|^{2}I_{\Rr}^{\mathcal{Z}}$, so the drive amplitude will reappear as the explicit factor $|\Nn|$ in the precision $\sigma_\Rr=1/\left(|\Nn|\sqrt{I_{\Rr}^{\mathcal{Z}}}\right)$.}
\begin{equation}
I_{\Rr}^{\mathcal{Z}}=\frac{|\partial_\Rr\mathcal{Z}|^{2}}{w}\,.
\label{eq:icplx}
\end{equation}
Profiling also confines the $\Rr$-dependence of the log-likelihood to a single term, which a common rescaling $z_j\to c\,z_j$ multiplies by $|c|^{2}$; the maximum-likelihood estimate of $\Rr$ is therefore a function of $z_\mathrm{K}/z_\mathrm{Rb}$ alone at any $\SNR$, the~amplitude setting only the sharpness of the likelihood. The~ratio is not, however, a~sufficient statistic away from the small-noise limit, since data with the same ratio but different amplitudes constrain $\Rr$ differently; it is in this sense that the complex ratio is the optimal observable, and~$\Dphi$ the optimal calibration-free~one.

The above result presumes the two channels' noise is dominated by \emph{independent}, per-channel readout noise. The~dominant \emph{correlated}
disturbance, the~magnetic background, requires a different treatment. We convert each channel's readout to an equivalent magnetic field by dividing by its own magnetic transfer function $T^B_j$, thus defining the \emph{field-referred}
data $x_j\equiv z_j/(\mathcal{K}_j T^B_j)$. In~these units the per-channel readout noise considered above reads $\tilde\sigma_j\equiv\sigma_j/(\mathcal{K}_j|T^B_j|)$. Because~the magnetic background is common to both channels, it adds a single fully correlated variance $S_B$ on top of the independent per-channel variances $\tilde\sigma_j^{2}$. Their sum $\tilde\sigma^{2}\equiv\tilde\sigma_\mathrm{K}^{2}+\tilde\sigma_\mathrm{Rb}^{2}$ is the readout variance of the differential combination $x_\mathrm{K}-x_\mathrm{Rb}$, in~which the common background cancels identically; it is the instrument's own noise, in~equivalent-field units, and~we will use it to normalize $S_B$. The~field-referred data (a two-dimensional vector) are then
\begin{equation}
x=\Nn s+(\text{noise})\,,
\label{eq:xmodel}
\end{equation}
with $s_j=T^b_j/T^B_j$ and the noise described by the covariance matrix
\begin{equation}
C=\begin{pmatrix}\tilde\sigma_\mathrm{K}^{2}+S_B & S_B\\[2pt] S_B & \tilde\sigma_\mathrm{Rb}^{2}+S_B\end{pmatrix}\,.
\label{eq:cov}
\end{equation}
The field-referred response inherits the affine form of Equation~\eqref{eq:affine}, $s_j(\theta)=a_j\cos\theta+b_j\sin\theta$, with~$a_j\equiv A_j/T^B_j$ and $b_j\equiv B_j/T^B_j$ the field-referred pure-proton and pure-neutron responses. Profiling the common drive $\Nn$ out of the two-channel likelihood gives the Fisher information
\begin{equation}
I_{\Rr}^{x}(S_B)=(\partial_\Rr s)^{\dagger}C^{-1}\partial_\Rr s-\frac{\big|s^{\dagger}C^{-1}\partial_\Rr s\big|^{2}}{s^{\dagger}C^{-1}s}\,.
\label{eq:gls}
\end{equation}
It is the $\Rr$-sensitivity of the signal weighted by the inverse noise $C^{-1}$, minus the part of that sensitivity parallel to the signal itself---a complex rescaling, which a change in $\Nn$ would mimic and profiling therefore removes. Carrying out the Lagrange reduction (Appendix~\ref{app:proof}) gives
\begin{equation}
I_{\Rr}^{x}(S_B)=\frac{|\partial_\Rr\mathcal{Z}|^{2}}{w_\mathrm{eff}},\qquad
w_\mathrm{eff}=w+S_B\left|\frac{1}{s_\mathrm{K}}-\frac{1}{s_\mathrm{Rb}}\right|^{2}\,.
\label{eq:weff}
\end{equation}
Adding the correlated background therefore only rescales the overall information about $\Rr$.\endnote{We neglect the spin-projection noise from the alkalis' mutual spin exchange. Correlated spin-projection noise predominantly excites the exchange-locked collective mode and therefore enters the two channels approximately common-mode, like the magnetic background itself: it is subsumed in $S_B$, a~fractional addition of only $\sim\!2\times10^{-4}$ for the $\sim\!0.1\,\mathrm{fT}/\sqrt{\mathrm{Hz}}$ budgeted in ref.~\cite{Letter}, while the residual non-common part lies well below the readout floors $\tilde\sigma_j$ across the band.}

\begin{figure}[!t]
\centering
\includegraphics[width=\textwidth]{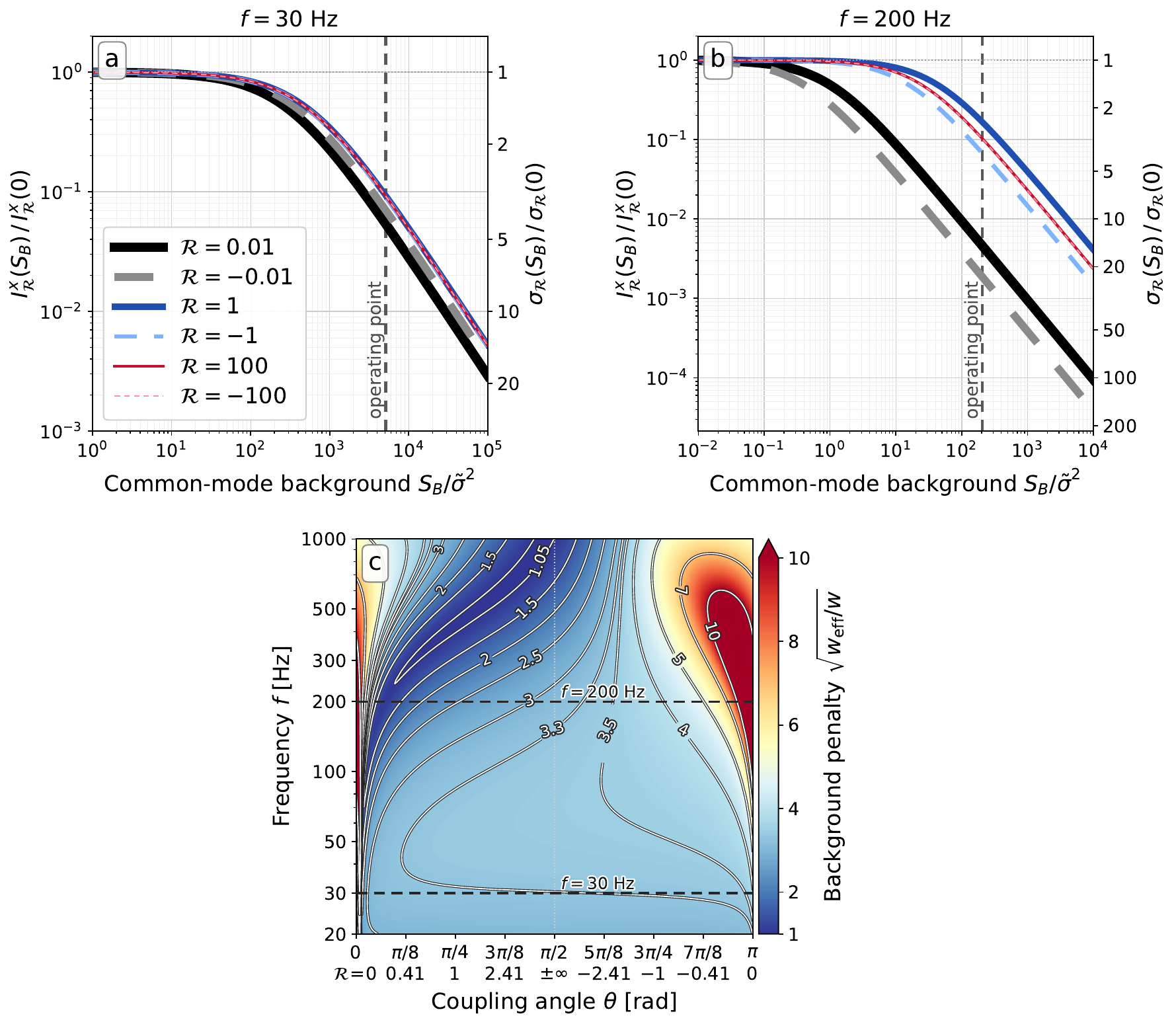}
\caption{The
 correlated magnetic background cost in coupling-ratio precision. (\textbf{a},\textbf{b})~The cost as a function of the background size at $30$ and $200$~Hz. The~left axis gives the Fisher information for $\Rr$ from Equation~\eqref{eq:gls}, normalized to the uncorrelated noise optimum, versus the relative common-mode magnetic background $S_B/\tilde\sigma^{2}$; the right axis gives the expected precision on $\Rr$, normalized the same way, $\sigma_\Rr(S_B)/\sigma_\Rr(0)$. Six coupling structures are shown, from~proton-dominated ($\Rr=\pm0.01$) through an equal mixture ($\Rr=\pm1$) to neutron-dominated ($\Rr=\pm100$); solid/dashed marks the sign of $\Rr$. The~vertical line marks the experiment's operating point at that frequency based on the numbers from ref.~\cite{Letter}. (\textbf{c})~The background penalty $\sqrt{w_\mathrm{eff}/w}=\sqrt{I_{\Rr}^{x}(0)/I_{\Rr}^{x}(S_B)}$ at the experiment's operating point as a function of the coupling angle $\theta$ ($\Rr=\tan\theta$; $\Rr$-value ticks accompany the $\theta$ ticks) and frequency. Constant-penalty contours are drawn in white; the dashed horizontal lines indicate the frequencies of panels~(\textbf{a},\textbf{b}).}
\label{fig:gls}
\end{figure}

Panels~(a,b) of Figure~\ref{fig:gls} show the effect of $S_B$ on the Fisher information and the resulting precision. The~operating point is taken from the noise values of ref.~\cite{Letter}: above $20$~Hz the correlated magnetic background is $6.2\,\mathrm{fT}/\sqrt{\mathrm{Hz}}$, well above the readout floor, whose field-referred value at $100$~Hz is $\tilde\sigma\approx\!0.22\,\mathrm{fT}/\sqrt{\mathrm{Hz}}$, so $S_B/\tilde\sigma^{2}=(6.2/0.22)^{2}\approx\!800$ there. That ratio is frequency dependent, since $\tilde\sigma_j\propto1/|T^B_j(f)|$ inherits the magnetic roll-off, a~factor of $\sim\!5$ between $30$ and $200$~Hz, placing the same background at $\approx\!5100$ at $30$~Hz and $\approx\!210$ at $200$~Hz: the vertical lines in the two panels. Panel~(c) of that figure maps the resulting penalty $\sqrt{w_\mathrm{eff}/w}$, read along this operating locus, across the whole plane of coupling angle and frequency. Over~most of the plane the penalty is $\approx\!3.3$, the~background costing about a factor of three in $\sigma_\Rr$, but~it climbs steeply in a narrow band of nearly pure-proton couplings, reaching $19$ for $\Rr=0$, $f=200$~Hz.

The penalty stays within a small factor even where the background exceeds the readout floor by orders of magnitude in power because the background is multiplied by the misalignment factor $M\equiv\left|1/s_\mathrm{K}-1/s_\mathrm{Rb}\right|^{2}$ from Equation~\eqref{eq:weff}. The~background displaces both channels alike, along $(1,1)$; the component of that displacement along the signal direction $s$ imitates the drive $\Nn$, which Equation~\eqref{eq:gls} already profiles away, so only the transverse remainder costs information---and $M$ measures exactly that.\endnote{A correlated disturbance entering the two channels with complex weights $c=(c_\mathrm{K},c_\mathrm{Rb})$ costs information only through its part not in the ratio $s_\mathrm{K}:s_\mathrm{Rb}$ (Appendix~\ref{app:proof}); the magnetic background, $c=(1,1)$, is the case carried by $M$ above. For~the \Rb-selective light-shift disturbance of ref.~\cite{Letter}, which departs from common mode by $|1-c_\mathrm{K}/c_\mathrm{Rb}|\approx0.07$ and whose common-mode part is already contained in $S_B$, the~remainder raises $\sigma_\Rr$ by at most $1.5\%$ at the level considered there.} Over most of the band the misalignment is small because both alkalis are largely controlled by the \emph{same} driven \He\ magnetization (Section~\ref{sec:review}), so $s_\mathrm{K}/s_\mathrm{Rb}$ is insensitive to the coupling structure and only the unequal Fermi-contact factors survive---a $\approx\!10\%$ fractional mismatch between the two responses at $100$~Hz. The~cancellation fails where the \He\ route does not dominate the dynamics: at high frequency, and~as $\Rr\to0$, since a proton coupling is barely driving the mostly neutron \He\ spin. This explains the high-penalty region in panel~(c).

\subsection{Phase and Amplitude Split the Information}

The information $I_{\Rr}^{x}(S_B)$ of Equation~\eqref{eq:weff} can be split into two terms via $|\partial_\Rr\mathcal{Z}|^{2}=(\mathrm{Re}\,\partial_\Rr\mathcal{Z})^{2}+(\mathrm{Im}\,\partial_\Rr\mathcal{Z})^{2}$, with~the real part the sensitivity of the amplitude ratio $\ln|T^b_\mathrm{K}/T^b_\mathrm{Rb}|$ and the imaginary part that of the differential phase $\Dphi$ of Equation~\eqref{eq:dphi}. In~the small-noise limit (in which the linearized quadratures of $\mathcal{Z}$ are independent Gaussians of common variance $w_\mathrm{eff}/|\Nn|^{2}$, the~phase does not wrap, and~the Cram\'er--Rao bound is attained),\endnote{The small quantity is $\varepsilon\equiv\sqrt{w_\mathrm{eff}}/|\Nn|$, the~standard deviation of each quadrature of $\mathcal{Z}$ and hence of the differential phase. Section~\ref{sec:results} delimits the small-noise regime, pairing the limit on $\varepsilon$ with a companion condition on the linearity of $\Dphi(\theta)$; the first is required for the phase/amplitude split of Equation~\eqref{eq:weff_split}, the~second only for attaining the Cram\'er--Rao bounds. Equation~\eqref{eq:weff} itself is the \emph{exact}
drive-profiled Fisher information of the model of Equation~\eqref{eq:model} at any $\SNR$, so neither the total information nor the intrinsic precision depends on either condition.\label{note:smallnoise}} the two terms are the information held separately by phase and amplitude (Appendix~\ref{app:proof}):\vspace{-3pt}

\begin{equation}
\begin{gathered}
I_{\Rr}^{x}(S_B)=I_{\Rr}^{\Dphi}(S_B)+I_{\Rr}^\mathrm{amp}(S_B),\\
I_{\Rr}^{\Dphi}(S_B)=\frac{(\mathrm{Im}\,\partial_\Rr\mathcal{Z})^{2}}{w_\mathrm{eff}}\,,\qquad
I_{\Rr}^\mathrm{amp}(S_B)=\frac{(\mathrm{Re}\,\partial_\Rr\mathcal{Z})^{2}}{w_\mathrm{eff}}\,.
\end{gathered}
\label{eq:weff_split}
\end{equation}

Both the differential phase and the amplitude ratio are invariant under the common drive $\Nn$, which cancels in any inter-channel ratio. They are \emph{not}, however, on~the same footing with respect to the \emph{instrument}. The~measured ratio is
\begin{equation}
\frac{z_\mathrm{K}}{z_\mathrm{Rb}}=\frac{\mathcal{K}_\mathrm{K}}{\mathcal{K}_\mathrm{Rb}}\,\frac{T^b_\mathrm{K}}{T^b_\mathrm{Rb}}+\text{noise}\,,
\end{equation}
in which the real, positive gain ratio $\mathcal{K}_\mathrm{K}/\mathcal{K}_\mathrm{Rb}$ multiplies the amplitude but leaves the phase untouched. The~differential phase $\Dphi$ is therefore immune to \emph{both} the common drive and the channel gains---genuinely calibration free---whereas the amplitude ratio is biased by any error in the gain ratio. In~this sense, $\Dphi$ is the calibration-free part of the~information.

\section{Results}
\label{sec:results}

Measurement precision is inversely proportional to the SNR, while the remaining factors that determine how well a quantity (here the coupling angle $\theta$) can be measured are encoded in what we call the \emph{intrinsic precision}
 $\kR$:
\begin{equation}
\sigma_\theta(S_B) = \frac{\kR(S_B)}{\SNR(S_B)}\,.
\label{eq:kappa}
\end{equation}
Apart from its dependence on the correlated magnetic background variance $S_B$, the~intrinsic precision is a function of the frequency and $\theta$. The~SNR is the matched-filter combination of the two field-referred channels,
\begin{equation}
\SNR^{2}(S_B)=|\Nn|^{2}\,s^{\dagger}C^{-1}s\,,
\label{eq:snr}
\end{equation}
with $s_j=T^b_j/T^B_j$ and $C=C(S_B)$ the covariance of Equation~\eqref{eq:cov}, reducing at $S_B=0$ to the sum of the squared single-channel signal-to-noise ratios,
\begin{equation}
\SNR^{2}(0)=|\Nn|^{2}\sum_j\left(\frac{\mathcal{K}_j|T^b_j|}{\sigma_j}\right)^{2}\,.
\end{equation}
The attainable precision in the measurement of $\theta$ is given by
\begin{equation}
\sigma_\theta(S_B)=\sigma_\Rr(S_B)\,\left|\frac{d\theta}{d\Rr}\right|\,,\qquad
\sigma_\Rr(S_B)=\frac{1}{|\Nn|\,\sqrt{I_{\Rr}^{x}(S_B)}}\,,
\label{eq:crb}
\end{equation}
where we used the Cram\'er--Rao bound, $I_{\Rr}^{x}(S_B)$ is given by Equation~\eqref{eq:weff}, and~$d\theta/d\Rr = \cos^2\theta$. The~unknown drive amplitude $|\Nn|$ cancels identically in $\kR$ since $\SNR\propto|\Nn|$ while $\sigma_\theta\propto1/|\Nn|$, so the intrinsic precision is a property of the transfer functions and the background~alone.

\begin{figure}[!t]
\centering
\includegraphics[width=\textwidth]{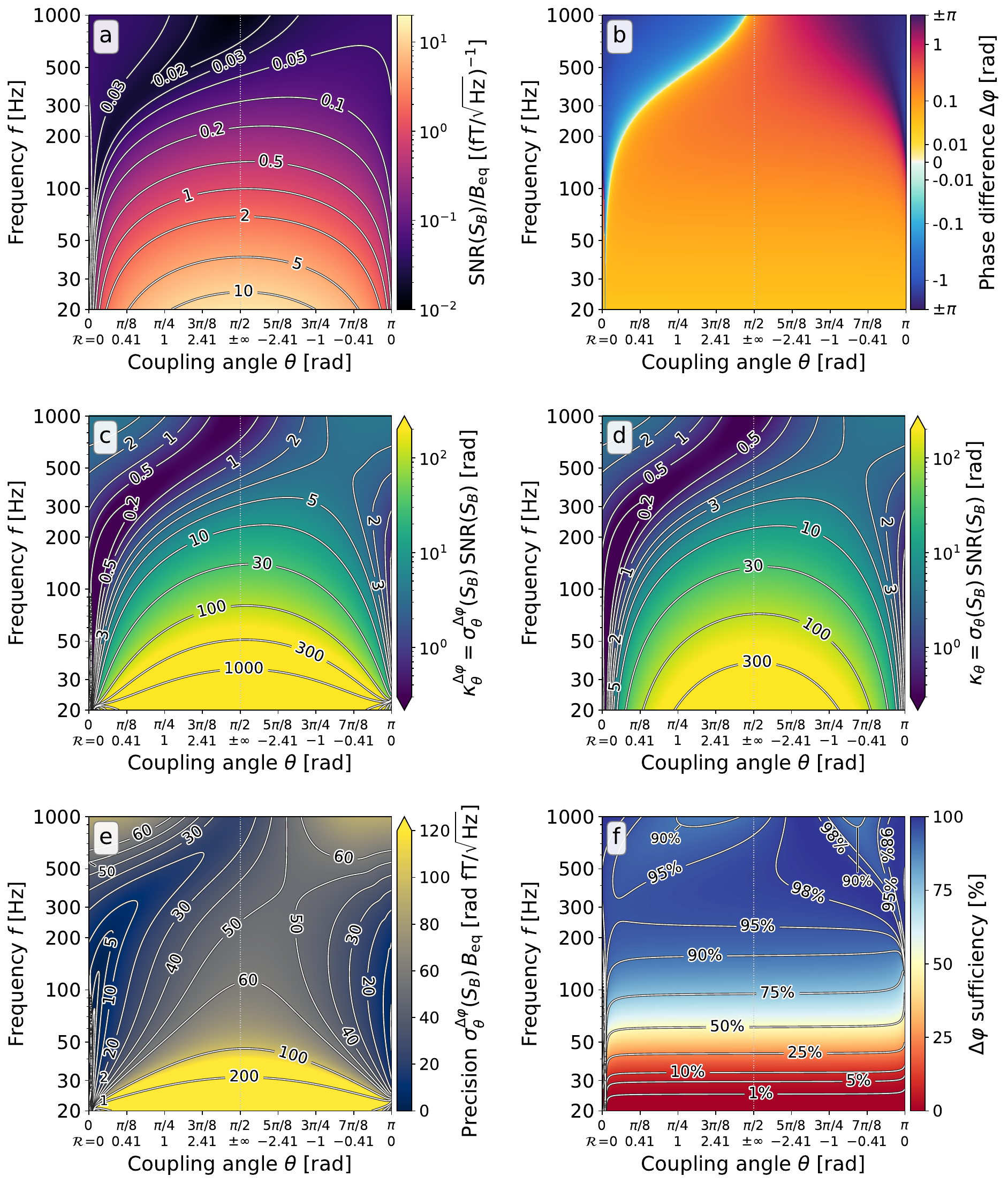}
\caption{Extraction
 of the coupling ratio $\Rr=\xi_n/\xi_p$. Each panel maps the $20$--$1000$~Hz band against the coupling angle $\theta$ ($\Rr=\tan\theta$; $\Rr$-value ticks accompany the $\theta$ ticks). (\textbf{a})~Achievable amplitude SNR per unit drive, the~drive referred to an equivalent field $B_\mathrm{eq}=|\Nn|/\gamma_e$: $\SNR(S_B)/B_\mathrm{eq}=\gamma_e\sqrt{s^{\dagger}C^{-1}s}$. (\textbf{b})~The calibration-free readout: the inter-species differential phase $\Dphi=\arg(T^b_\mathrm{K}/T^b_{\mathrm{Rb}})$ (logarithmic color scale, linear for $|\Dphi|<10^{-2}\,\mathrm{rad}$). (\textbf{c})~Intrinsic precision provided by that readout, $\kDphi=\sigma_\theta^{\Dphi}(S_B)\cdot\SNR(S_B)$, with~$\sigma_\theta^{\Dphi}$ the Cram\'er--Rao bound from the differential phase alone. (\textbf{d})~Intrinsic precision provided by the full two-channel measurement, $\kR=\sigma_\theta(S_B)\cdot\SNR(S_B)$. (\textbf{e})~Achievable precision on the coupling angle from the differential phase alone, per unit drive: $\sigma_\theta^{\Dphi}(S_B)\,B_\mathrm{eq}=1/(\gamma_e\sqrt{I_\theta^{\Dphi}(S_B)})$ with $I_\theta^{\Dphi}(S_B)=I_{\Rr}^{\Dphi}(S_B)/\cos^{4}\theta$, the~ratio of panels~(\textbf{c}) and~(\textbf{a}). (\textbf{f})~Differential-phase sufficiency $I_{\theta}^{\Dphi}(S_B)/I_{\theta}^{x}(S_B)$, the~percentage of the variance-optimal complex-ratio information retained by $\Dphi$ alone. The~per-unit-drive maps of panels~(\textbf{a},\textbf{e}) are referred to a $1$~Hz band. Panels~(\textbf{a},\textbf{c}--\textbf{e}) are evaluated at the operating-point background; panels~(\textbf{b},\textbf{f}) are fixed by the transfer functions alone. Panels~(\textbf{c},\textbf{e},\textbf{f}) rest on the phase/amplitude split of Equation~\eqref{eq:weff_split} and inherit its small-noise restriction, while panels~(\textbf{a},\textbf{b},\textbf{d}) do~not.}
\label{fig:Robs}
\end{figure}

Panel~(a) of Figure~\ref{fig:Robs} maps the achievable SNR per unit drive at the operating background, $\SNR(S_B)/B_\mathrm{eq}=\gamma_e\sqrt{s^{\dagger}C^{-1}s}$, the~drive being referred to an equivalent field $B_\mathrm{eq}=|\Nn|/\gamma_e$. It is largest for neutron-dominated couplings across most of the band and falls by up to a factor of $\sim\!30$ toward $\theta=0,\pi$, a~deficit deepest at low frequency. The~response strength reflects a competition between two drive routes of opposite and frequency-independent preference. The~\He spin is essentially the neutron's, so the \He-mediated route favors neutron couplings by a factor of $32$. The~proton spin fractions of \K and \Rb carry opposite signs, so the direct alkali route drives the two channels differentially. The~inverse-noise weighting suppresses the common mode, in~which the near-common-mode neutron drive largely sits, and~leaves the differential component at full weight, favoring proton couplings by $4.7$ at $200$~Hz, a~preference flat to $2\%$ across the band. At~the bottom of the band the \He route carries the alkali signal almost alone and the deficit approaches its full factor of $32$; as the \He share falls with frequency, the alkali route's proton preference takes over near $450$~Hz, above~which proton-dominated couplings reach the larger~SNR.

Panel~(b) maps the calibration-free readout, the~inter-species differential phase $\Dphi=\arg(T^b_\mathrm{K}/T^b_{\mathrm{Rb}})$. Panels~(c,d) show the intrinsic precision provided by that readout, $\kDphi=\sigma_\theta^{\Dphi}(S_B)\cdot\SNR(S_B)$, and~the intrinsic precision $\kR$ provided by the full two-channel measurement, with~$I_\theta^{x}(S_B)=I_{\Rr}^{x}(S_B)/\cos^{4}\theta$. The~two are nearly indistinguishable above $\sim\!150$~Hz; below that, $\kDphi$ diverges in places where $\partial_\theta\Dphi$ vanishes and only the amplitude ratio still responds to the coupling. They share their coupling dependence: below $\sim\!450$~Hz proton-dominated couplings have better intrinsic precision, needing $21$ times less SNR at $200$~Hz than neutron-dominated ones for the same precision on $\theta$, and~the ordering reverses above that frequency. Since $s_j(\theta)=a_j\cos\theta+b_j\sin\theta$, varying $\theta$ only rotates the pair $(s,\partial_\theta s)$ within the fixed field-referred proton and neutron responses $a$ and $b$, which leaves unchanged the area $\mathcal{A}$ that pair spans in the inverse-noise metric $C^{-1}$, a~function of the frequency alone. The~intrinsic precision then depends on the response strength as
\begin{equation}
\kR=\frac{s^{\dagger}C^{-1}s}{\mathcal{A}}=\frac{1}{\gamma_e^{2}\mathcal{A}}\left[\frac{\SNR(S_B)}{B_\mathrm{eq}}\right]^{2}\,,\qquad\mathcal{A}=\frac{|a_\mathrm{K}b_\mathrm{Rb}-a_\mathrm{Rb}b_\mathrm{K}|}{\sqrt{\det C}}\,.
\label{eq:kappa_area}
\end{equation}
Therefore, at~any given frequency, the~couplings with the strongest response have the worst intrinsic~precision.

Panel~(e) presents the attainable precision on $\theta$ from the $\Dphi$ measurement, the~ratio of panels~(c) and~(a), $\sigma_\theta^{\Dphi}(S_B)\,B_\mathrm{eq}=\kDphi/[\SNR(S_B)/B_\mathrm{eq}]$. Panel~(f) quantifies how much of the full coupling-ratio information the calibration-free $\Dphi$ alone captures, through the \emph{differential-phase sufficiency}
$I_{\theta}^{\Dphi}(S_B)/I_{\theta}^{x}(S_B)=\sin^{2}(\arg\partial_\theta\mathcal{Z})$.

Above $\sim\!100$~Hz, $\Dphi$ is near-sufficient over most of the coupling range. At~lower frequencies, the sufficiency degrades sharply---the retained fraction falls below $25\%$ near $40$~Hz and recovering it would sharpen the precision on $\theta$ by the factor $\big[I_{\theta}^{x}(S_B)/I_{\theta}^{\Dphi}(S_B)\big]^{1/2}\gtrsim2$, and~at the bottom of the band $\Dphi$ retains only about one percent of the information. The~two regimes differ in what drives the channels. Below~$\sim\!100$~Hz, both are driven predominantly through the same quasi-static \He magnetization and respond nearly in phase, so a change in the coupling is expressed almost entirely in the ratio of the two amplitudes. At~higher frequencies, the~\He roll-off and the alkali antiresonances of Figure~\ref{fig:dphicuts} separate the two channels' phase lags, and~the coupling sweeps the differential phase~instead.

Panels~(c,e,f) are valid in the small-noise regime of Note~8. Monte Carlo simulation on proton-dominated couplings shows that the full profiled maximum-likelihood estimator and the one based on $\Dphi$ alone attain their Cram\'er--Rao bounds to within $10\%$ when $\varepsilon\le0.3$ and the slope $\partial_\theta\Dphi$ changes by less than $30\%$ over one standard deviation of the $\Dphi$-based estimate of $\theta$. These conditions set a floor on the detection $\SNR$. Above~$\sim\!150$~Hz the floor is typically of order ten, and~$\approx80$ on the proton-dominated couplings of best intrinsic precision at $200$~Hz, an~$\SNR$ at which already $\sigma_\theta\lesssim0.01$~rad. Toward lower frequencies it grows, to~$\approx330$ at the \Rb antiresonance at $\Rr=0$ and, above~$100$~Hz, higher only near a null of the \Rb exotic response at $90$~Hz and $\Rr\approx0.016$. Below $100$~Hz it typically exceeds $100$ and diverges at that null and where $\partial_\theta\Dphi$ vanishes, as~$\kDphi$ does. At~a detection $\SNR$ below the floor, panels~(c) and~(e) give Cram\'er--Rao bounds rather than attained~precisions.

The frequencies quoted in this section are specific to the operating point of Table~\ref{tab:params}. Varying $\alpha$ over $0.46$--$0.69$ ($\pm20\%$ around the baseline $0.578$) shifts the \K antiresonance between $195$ and $111$~Hz, the~$\sim\!100$~Hz near-sufficiency threshold between $126$ and $67$~Hz, and~the $\sim\!450$~Hz reversal between $467$ and $437$~Hz; the two-band structure of Figure~\ref{fig:Robs} persists~throughout.

\section{Operating-Point and Instrumental Tolerances, and~Their~Control}
\label{sec:systematics}

The previous sections assume that the transfer functions are known and that the measured differential phase is free of instrumental offsets. We define the tolerance on each systematic effect as the size at which the error $\delta\Dphi$ it causes in the differential phase equals the statistical uncertainty $\sigma_{\Dphi}$ of that phase, the~quantity $\varepsilon$ of Note~8. By~Equation~\eqref{eq:snr}, $\sigma_{\Dphi}=\sqrt{w_\mathrm{eff}\,s^{\dagger}C^{-1}s}\big/\SNR$, so every tolerance scales as $1/\SNR$. We quote each at a reference $\SNR=100$, comparable with the small-noise floor on the couplings of best intrinsic precision at $200$~Hz, as~the tightest value over the frequency--coupling plane unless a narrower domain is stated. Where a tolerance is tightest, $\sigma_{\Dphi}$ is at the milliradian level; elsewhere at $\SNR=100$ it reaches $0.23$~rad on those couplings and, outside the small-noise regime, $\approx\!1$~rad at the $\Rr=0$ \Rb antiresonance, where the tolerances are correspondingly~looser.

\subsection{Operating~Point}
\label{sec:sys-oppoint}

The antiresonance frequencies depend on the longitudinal compensation field $B_z$ as strongly as on $\Rr$. A~$0.5\%$ change in $B_z$, $\approx\!1$~nT at the operating point of Table~\ref{tab:params}, mimics $\delta\Rr=0.01$ (Figure~\ref{fig:dphicuts}), so a drift of the operating point would be misread as a change in the coupling ratio. The~two dependences can nevertheless be separated. Since $\Rr$ enters the Bloch equations only through the exotic couplings of Equation~\eqref{eq:xij}, the~magnetic transfer functions $T^B_j$ do not depend on $\Rr$, and~the magnetic response determines the operating point in~situ, independently of any candidate signal. The~compensation field must be held to within $4\times10^{-4}$ of its value, $\approx\!80$~pT, at~the reference $\SNR$.

The limiting drift is expected to be that of the compensation point, which follows the slowly varying \He magnetization. It can be corrected by re-setting $B_z$ against the atoms, continuously~\cite{Klinger:2022tfp} or periodically~\cite{Padniuk2024thesis,Lu2020}. A~transverse field is briefly modulated (for $12$~s in ref.~\cite{Padniuk2024thesis}), and~the alkali response to it, recorded as a function of $B_z$, serves as the error signal. This response is a dispersive resonance~\cite{Klinger:2022tfp}; ref.~\cite{Lu2020} locates its zero crossing to $0.05$~nT within a minute. The~zero crossing, which marks the \He field, differs from the compensation point by a set-point offset, so $B_z$ is re-set to it plus that offset~\cite{Padniuk2024thesis,Lu2020}. In~a dual-alkali cell the modulation frequency must lie well above the \He Larmor frequency, $6.6$~Hz, so that the \He barely follows it, and~well below the alkali--alkali spin-exchange rate, $\approx\!5\times10^{4}$~Hz, so that the alkalis remain locked and present a single zero crossing for the pair. A~frequency of a few hundred hertz satisfies both~conditions.

The requirement then concerns the drift of $B_z$ between re-settings rather than its absolute stability. If~the \He magnetization drifts at the rate of the response drifts reported in ref.~\cite{Padniuk2024thesis}, $\le\!0.02\%$ per hour, the~\He field moves by $\approx\!40$~pT per hour, a~drift that the hourly re-setting practiced there keeps within the tolerance at the reference $\SNR$; at higher $\SNR$ the interval shortens as $1/\SNR$.

At this operating point the model of Section~\ref{sec:review} puts the set-point offset at $\approx\!3$~nT, of~which $2$~nT come from the alkali magnetization terms of Equation~(6) of ref.~\cite{Letter} and the rest from the partial response of the \He to the modulation. Refs.~\cite{Klinger:2022tfp,Padniuk2024thesis,Lu2020} could neglect this offset or fix it once. Here it exceeds the tolerance by more than an order of magnitude and must be known to a relative accuracy of $\approx\!2.5\%$. It follows mainly the alkali magnetization, which, to~judge from the response drifts of ref.~\cite{Padniuk2024thesis}, changes by only a few percent over $20$~days, so a daily re-calibration~suffices.

The offset is calibrated with the synthetic field of ref.~\cite{Rosenzweig2024}, a~coil field combined with light-induced fictitious fields on individual alkalis. Applied far below the detection band, it also yields the light-shift difference between the two alkalis, which a single setting of $B_z$ cannot cancel. Since a fractional offset of $10^{-5}$ in $B_z$ rotates the differential phase of the magnetic response at $0.05$~Hz by $63$~mrad, this low-frequency measurement resolves the compensation point of each channel well within the tolerance. With~an hourly re-setting against the \He field and a daily calibration of the offset, $B_z$ stays within its tolerance, and~a drift of the operating point cannot be mistaken for a change in the coupling~ratio.

\subsection{Model~Parameters}
\label{sec:sys-params}

The remaining parameters of Table~\ref{tab:params} bias $\Rr$ in the same way, but~more weakly. Their tolerances at the reference $\SNR$ range from $2\times10^{-3}$ on the \He density and polarization to $2\times10^{-2}$ on the \K density, all at least five times looser than that on $B_z$. They presume that $B_z$ is re-tuned to the compensation point after any change of these parameters, as~the re-setting against the \He field ensures. The~same hourly re-setting meets the tightest of them, on~the \He magnetization, since it locates the \He field to a few parts in $10^{4}$~\cite{Lu2020}. The~alkali densities and relaxation rates follow the cell temperature at $\approx\!5\%$ per kelvin, and~comagnetometer ovens hold the temperature to $\lesssim\!0.1\,^{\circ}$C~\cite{Kornack2005thesis,Vasilakis2011thesis}, which corresponds to half a percent in density; the percent-level tolerances depend on the oven, and~any tighter ones on~re-calibration.

Swept across frequency, the~synthetic field re-determines these parameters and the alkali polarizations. Its fictitious fields also reach the differential alkali mode, which a magnetic field barely excites. With~response drifts of $\le\!0.02\%$ per hour~\cite{Padniuk2024thesis}, the~daily sweep that accompanies the offset calibration covers the \K density with a fourfold margin. The~only combination the sweep leaves undetermined, $n_\mathrm{He}$ and $P_z^\mathrm{He}$ separately at fixed magnetization, is one to which the inferred $\Rr$ is also~insensitive.

\subsection{Probe~Cross-Talk}
\label{sec:sys-crosstalk}

Because each probe also senses the dispersive tail of the other alkali, each channel contains a fraction $\chi$ of the other channel~\cite{Letter}. The~effect could be enhanced near the antiresonances, where one alkali's response is small while the other's is not. To~first order, the~admixture shifts the log-ratio $\mathcal{Z}$ by $\delta\mathcal{Z}=\chi\,(T^b_\mathrm{Rb}/T^b_\mathrm{K}-T^b_\mathrm{K}/T^b_\mathrm{Rb})$. Equal fractions in the two channels are the worst case, because~the two admixtures then enter $\mathrm{Im}\,\delta\mathcal{Z}$ with the same sign. At~$\Rr=0$ the antiresonances are shallow, with~$0.6\lesssim|T^b_\mathrm{K}/T^b_\mathrm{Rb}|\lesssim2.6$ across the band, and~at a deep antiresonance the shift and $\sqrt{w_\mathrm{eff}}$ grow at the same rate, leaving the tolerance unchanged. The~resulting tolerance at the reference $\SNR$ is $\chi\lesssim10^{-2}$, tightest at the lower band~edge.

The admixture need not be measured, since it is fixed by the probe detunings and the alkali density ratio through Equation~(1) of ref.~\cite{Letter}. For~the probes of that work it is $\approx\!-0.8\%$ on the \K channel and $\approx\!-0.1\%$ on the \Rb channel, the~percent level stated there. Once these values are entered in the model, only their uncertainty biases $\Rr$; a $10\%$ error in the density ratio leaves a residual admixture of $10^{-3}$, an~order of magnitude below the tolerance. Cross-talk is therefore negligible, near~the antiresonances or elsewhere, except~in the immediate vicinity of the null of the \Rb exotic response identified in Section~\ref{sec:results}, where $|T^b_\mathrm{K}/T^b_\mathrm{Rb}|$ exceeds $100$. There, at~$\chi=10^{-2}$, the~admixture exceeds the \Rb response it contaminates, the~expansion in $\chi$ fails, and~the null is not~observable.

\subsection{Differential Chain~Phase}
\label{sec:sys-chain}

The calibration-free property presumes real, positive gains $\mathcal{K}_j$. A~phase difference $\varphi_d\equiv\arg(\mathcal{K}_\mathrm{K}/\mathcal{K}_\mathrm{Rb})$ between the two detection chains adds to $\Dphi$ and has the same tolerance, $\sigma_{\Dphi}$. At~the reference $\SNR$ this is $0.5$~mrad at the most demanding point, the~narrow near-common-mode trough of panel~(a) of Figure~\ref{fig:Robs} at $\Rr\approx\!0.06$ near $80$~Hz, and~$3.7$~mrad at $\Rr=1$ over $100$--$200$~Hz.

The static part of $\varphi_d$, of~order $0.1$~mrad at $80$~Hz for photodiode amplifiers of $100$~kHz bandwidth matched to $10\%$, is measured and removed by the calibration described below; the shared lock-in of ref.~\cite{Letter} contributes nothing to it. The~tolerance therefore constrains only the varying part, namely the drift of the electronics between calibrations and the dependence of each chain's phase $\varphi_j$ on its signal amplitude $|z_j|$ at the demodulator. For~a conversion coefficient $\zeta_j\equiv\partial\varphi_j/\partial\ln|z_j|$ with $|\zeta_j|\lesssim0.05$~rad, the~$0.5$~mrad bound is met by holding the probe power of each channel, which scales $|z_j|$, stable to $0.5\%$, a~level that standard intensity stabilization provides~\cite{Lu2020,Kornack2005thesis,Vasilakis2011thesis}.

The sweep of Section~\ref{sec:sys-params}, which necessarily spans the band, measures $\varphi_d$ in~situ against the magnetic response at each frequency. Applied at several amplitudes spanning those of a detectable signal, with~each probe's power scaled in turn, it also measures $\zeta_j$. The~differential phase of the magnetic response, against~which $\varphi_d$ is measured, can be computed in advance and is insensitive to residual errors in the operating point and parameters; with these at their tolerances, it moves by less than $0.25$~mrad across the band, which is below the tolerance on $\varphi_d$. Only the value of $\varphi_d$ at the signal frequency enters $\Dphi$, and~a candidate is followed by a dedicated measurement at its frequency. In~that measurement, the~synthetic field, set to mimic the exotic drive for trial values of $\Rr$, measures $\Dphi(\Rr)$ directly, including the chain~phase.

\section{Summary and~Conclusions}
\label{sec:conclusions}

Ref.~\cite{Letter} proposed the inter-species phase difference $\Dphi$ of a single-cell dual-alkali comagnetometer as a calibration-free readout of the coupling ratio $\Rr=\xi_n/\xi_p$, which the two alkalis encode in their \emph{relative}
response because they carry different mixtures of the neutron and proton~couplings.

Treating the extraction of $\Rr$ as a statistical estimation problem, we have shown here that profiling the common drive out of the two-channel likelihood leaves the maximum-likelihood estimate of $\Rr$ depending on the two channels only through their complex ratio, with~Fisher information $|\partial_\Rr\mathcal{Z}|^2/w_\mathrm{eff}$ at any $\SNR$, that $\mathcal{Z}$ splits into $\Dphi$ and an amplitude ratio, and~that only $\Dphi$ is immune to the inter-channel gain. The~measured ratio itself carries the full information only in the small-noise~regime.

We mapped, over~the $20$--$1000$~Hz band and the full range of coupling angles, the~differential phase together with the quantities that govern its use: the attainable detection SNR per unit drive, the~intrinsic precision on the coupling angle provided by $\Dphi$ and by the full two-channel measurement, the~overall precision that $\Dphi$ affords at a given drive, and~the share of the coupling information it retains. At~any given frequency the couplings with the strongest response have the worst intrinsic precision. Below~$\sim\!450$~Hz the neutron-dominated couplings reach the larger SNR while the proton-dominated ones have better intrinsic precision, and~above it both statements reverse. We also found that above $\sim\!100$~Hz, $\Dphi$ is near-sufficient over most of the coupling range and is therefore the natural and robust readout for the coupling ratio. At~lower frequencies the amplitude ratio holds a substantial part of the information, recoverable only at the price of a calibrated inter-channel gain. Finally, we derived the tolerances on the operating point and on the instrumental effects that could bias $\Dphi$, and~showed that they can be met in~situ.

The same $\mathcal{Z}$ ratio construction that removes the unknown drive also removes any information about the size of the signal, so the extraction of $\Rr$ and the detection of the field are complementary uses of the same two channels, the~latter resting on the amplitude combination that cancels the magnetic background~\cite{Letter}. Within~the structure established here, no more elaborate observable can improve upon $\Dphi$ without forfeiting the calibration-free property, the~remaining information residing entirely in the amplitude ratio. A~single-cell dual-alkali comagnetometer thus offers, beyond~its sensitivity as a field detector, an~intrinsic calibration-free handle on the \emph{internal structure}
of a candidate dark matter~coupling.

\vspace{6pt}

\authorcontributions{Conceptualization, Y.R., Y.K. and~E.S.; methodology, Y.R. and Y.K.; software, Y.R., Y.K. and Y.J.; validation, Y.R. and Y.K.; formal analysis, Y.R., Y.K. and Y.J.; writing---original draft preparation, Y.K.; writing---review and editing, Y.R., E.S., Y.J. and R.F.; visualization, Y.K.; supervision, Y.K., E.S. and R.F.; project administration, Y.R., Y.K., E.S. and R.F.; funding acquisition, Y.K., E.S., M.G. and R.F. All authors have read and agreed to the published version of the manuscript.}

\funding{We
 gratefully acknowledge support from the Gordon and Betty Moore Foundation, Simons Foundation, Alfred P. Sloan Foundation, and~John Templeton Foundation. This work was supported in part also by the United States--Israel Binational Science Foundation Grants No.\ 2016635 and No.\ 2018257 and the Israel Innovation Authority Grants No.\ 67082 and No.\ 74482. YK is supported in part by the Israel Science Foundation Grant No.~1666/22.}

\dataavailability{The data presented in this study are available on request from the corresponding author.}

\acknowledgments{During the preparation of this manuscript, the~authors used Anthropic's Claude models (Opus~4.8, Opus~5, and~Fable~5) for the purposes of text drafting and editing, figure generation, and~numerical analysis and cross-checking. The~authors have reviewed and edited the output and take full responsibility for the content of this~publication.}

\conflictsofinterest{The authors declare no conflicts of~interest.}

\appendixtitles{yes}
\appendixstart
\appendix
\section{Reduction of the Profiled Information and the Phase/Amplitude~Split}
\label{app:proof}

This appendix proves that the profiled two-channel Fisher information of Equation~\eqref{eq:gls} reduces to the closed form of Equation~\eqref{eq:weff}, and~that in the small-noise limit its two quadratures carry the phase/amplitude split of Equation~\eqref{eq:weff_split}. Throughout, $s=(s_\mathrm{K},s_\mathrm{Rb})^{\top}$ with $s_j=T^b_j/T^B_j$ is the field-referred response of Section~\ref{sec:facts}, $\partial_\Rr s$ its coupling-ratio derivative, and~$C$ the covariance of Equation~\eqref{eq:cov}; we abbreviate the inverse-noise inner product $\langle p,q\rangle\equiv p^{\dagger}C^{-1}q$. In~this notation, Equation~\eqref{eq:gls} is the ratio of the Gram determinant of $s$ and $\partial_\Rr s$ to the squared norm $\langle s,s\rangle$:
\begin{equation}
I_{\Rr}^{x}(S_B)=\langle\partial_\Rr s,\partial_\Rr s\rangle-\frac{|\langle s,\partial_\Rr s\rangle|^{2}}{\langle s,s\rangle}=\frac{\det G}{\langle s,s\rangle},\qquad
G=\begin{pmatrix}\langle s,s\rangle & \langle s,\partial_\Rr s\rangle\\[2pt]\langle\partial_\Rr s,s\rangle & \langle\partial_\Rr s,\partial_\Rr s\rangle\end{pmatrix}\,.
\label{eq:app-gram}
\end{equation}

Let $J=[\,s\mid\partial_\Rr s\,]$, so that $G=J^{\dagger}C^{-1}J$. By~multiplicativity of the determinant, $\det G=\overline{\det J}\,\det(C^{-1})\,\det J=|\det J|^{2}/\det C$; what remains is to evaluate $\det J$ and $\langle s,s\rangle$.

The determinant of $J$ is
\begin{equation}
\det J=s_\mathrm{K}\,\partial_\Rr s_\mathrm{Rb}-s_\mathrm{Rb}\,\partial_\Rr s_\mathrm{K}=-\,s_\mathrm{K}s_\mathrm{Rb}\left(\frac{\partial_\Rr s_\mathrm{K}}{s_\mathrm{K}}-\frac{\partial_\Rr s_\mathrm{Rb}}{s_\mathrm{Rb}}\right)=-\,s_\mathrm{K}s_\mathrm{Rb}\,\partial_\Rr\mathcal{Z}\,,
\label{eq:app-wronskian}
\end{equation}
where $\partial_\Rr\mathcal{Z}=\partial_\Rr\ln(s_\mathrm{K}/s_\mathrm{Rb})=\partial_\Rr\ln(T^b_\mathrm{K}/T^b_\mathrm{Rb})$, the~two logarithms differing only by the $\Rr$-independent $\ln(T^B_\mathrm{K}/T^B_\mathrm{Rb})$. Hence $|\det J|^{2}=|s_\mathrm{K}s_\mathrm{Rb}|^{2}\,|\partial_\Rr\mathcal{Z}|^{2}$.

Since $(\det C)\,C^{-1}=\operatorname{adj}C$, $\det C\,\langle s,s\rangle=s^{\dagger}(\operatorname{adj}C)\,s$. For~the real symmetric $C$ of Equation~\eqref{eq:cov}, $\operatorname{adj}C=\left(\begin{smallmatrix}\tilde\sigma_\mathrm{Rb}^{2}+S_B & -S_B\\ -S_B & \tilde\sigma_\mathrm{K}^{2}+S_B\end{smallmatrix}\right)$, and~\begin{equation}
s^{\dagger}(\operatorname{adj}C)\,s=\tilde\sigma_\mathrm{Rb}^{2}|s_\mathrm{K}|^{2}+\tilde\sigma_\mathrm{K}^{2}|s_\mathrm{Rb}|^{2}+S_B\,|s_\mathrm{K}-s_\mathrm{Rb}|^{2}\,.
\label{eq:app-adj}
\end{equation}
Dividing by $|s_\mathrm{K}s_\mathrm{Rb}|^{2}$,
\begin{equation}
\frac{s^{\dagger}(\operatorname{adj}C)\,s}{|s_\mathrm{K}s_\mathrm{Rb}|^{2}}=\underbrace{\frac{\tilde\sigma_\mathrm{K}^{2}}{|s_\mathrm{K}|^{2}}+\frac{\tilde\sigma_\mathrm{Rb}^{2}}{|s_\mathrm{Rb}|^{2}}}_{w}+S_B\,\left|\frac{1}{s_\mathrm{K}}-\frac{1}{s_\mathrm{Rb}}\right|^{2}=w_\mathrm{eff}\,,
\label{eq:app-weff}
\end{equation}
the first two terms being $w$ of Equation~\eqref{eq:weight} (in field-referred form, $\tilde\sigma_j/|s_j|=\sigma_j/(\mathcal{K}_j|T^b_j|)$).

Combining Equations~\eqref{eq:app-gram}--\eqref{eq:app-weff} and using $\det C\,\langle s,s\rangle=s^{\dagger}(\operatorname{adj}C)\,s$, we find that the common factor $|s_\mathrm{K}s_\mathrm{Rb}|^{2}$ cancels:
\begin{equation}
I_{\Rr}^{x}(S_B)=\frac{\det G}{\langle s,s\rangle}=\frac{|\det J|^{2}}{s^{\dagger}(\operatorname{adj}C)\,s}=\frac{|\partial_\Rr\mathcal{Z}|^{2}}{w_\mathrm{eff}}\,,
\label{eq:app-result}
\end{equation}
which is Equation~\eqref{eq:weff}. At~$S_B=0$, $w_\mathrm{eff}=w$ and Equation~\eqref{eq:app-result} reduces to Equation~\eqref{eq:icplx}; the Lagrange identity quoted in Note~4 in Section~\ref{sec:facts} is the special case $C=\operatorname{diag}(\tilde\sigma_\mathrm{K}^2,\tilde\sigma_\mathrm{Rb}^2)$.

The reduction above uses the specific form of Equation~\eqref{eq:cov} only in Equation~\eqref{eq:app-adj}. For~any Hermitian $C$, $w_\mathrm{eff}=s^{\dagger}(\operatorname{adj}C)\,s/|s_\mathrm{K}s_\mathrm{Rb}|^{2}$, so a disturbance of density $S$ entering the channels with complex weights $(c_\mathrm{K},c_\mathrm{Rb})$ adds $S\,|c_\mathrm{K}/s_\mathrm{K}-c_\mathrm{Rb}/s_\mathrm{Rb}|^{2}$ to $w_\mathrm{eff}$, of~which Equation~\eqref{eq:weff} is the case $c=(1,1)$. Only the part of such a disturbance that is not proportional to the signal direction $s$ costs~information.

The split of Equation~\eqref{eq:weff_split} follows from the same $w_\mathrm{eff}$, read now as a variance. Linearizing the field-referred log-ratio $\ln(x_\mathrm{K}/x_\mathrm{Rb})$ about its noiseless value $\ln(s_\mathrm{K}/s_\mathrm{Rb})$ gives a fluctuation $\delta\mathcal{Z}$ with $\Nn\,\delta\mathcal{Z}=u^{\top}(x-\Nn s)$, where $x-\Nn s$ is the field-referred noise of Equation~\eqref{eq:xmodel}, of~covariance $C$, and~$u=(1/s_\mathrm{K},\,-1/s_\mathrm{Rb})^{\top}$. Being a fixed linear combination of jointly circular Gaussians, $\Nn\,\delta\mathcal{Z}$ is itself circular, so its two quadratures are independent Gaussians of the common variance (formed with $C$ itself, not the inverse-noise metric)
\begin{equation}
u^{\dagger}Cu=\frac{\tilde\sigma_\mathrm{K}^{2}}{|s_\mathrm{K}|^{2}}+\frac{\tilde\sigma_\mathrm{Rb}^{2}}{|s_\mathrm{Rb}|^{2}}+S_B\left|\frac{1}{s_\mathrm{K}}-\frac{1}{s_\mathrm{Rb}}\right|^{2}=w_\mathrm{eff}\,,
\label{eq:app-var}
\end{equation}
which is Equation~\eqref{eq:app-weff} once more, reached as a variance rather than as an algebraic combination. The~two quadratures are therefore independent Gaussian observations whose means carry the $\Rr$-dependence of $\mathrm{Re}\,\mathcal{Z}=\ln|T^b_\mathrm{K}/T^b_\mathrm{Rb}|$ and $\mathrm{Im}\,\mathcal{Z}=\Dphi$, the~two log-ratios differing only by the $\Rr$-independent constant noted below Equation~\eqref{eq:app-wronskian}. Their common variance $w_\mathrm{eff}/|\Nn|^{2}$ depends on $\Rr$ as well, but~only through an overall scale; that scale is degenerate with the profiled drive amplitude and therefore carries no information about $\Rr$. The~Fisher information of such an observation is the squared derivative of its mean divided by that variance, giving $(\mathrm{Re}\,\partial_\Rr\mathcal{Z})^{2}/w_\mathrm{eff}$ and $(\mathrm{Im}\,\partial_\Rr\mathcal{Z})^{2}/w_\mathrm{eff}$ per unit $|\Nn|^{2}$---the two terms of Equation~\eqref{eq:weff_split}---and independent observations contribute additively. Their sum $|\partial_\Rr\mathcal{Z}|^{2}/w_\mathrm{eff}$ is Equation~\eqref{eq:app-result}, so the linearized quadratures recover the exact profiled information. The~ratio $I_{\Rr}^{\Dphi}(S_B)/I_{\Rr}^\mathrm{amp}(S_B)=\tan^{2}(\arg\partial_\Rr\mathcal{Z})$ is independent of $w_\mathrm{eff}$: the background $S_B$ rescales the total information but leaves the phase/amplitude division untouched, and~with it the differential-phase sufficiency $I_{\Rr}^{\Dphi}(S_B)/I_{\Rr}^{x}(S_B)$.

\printendnotes[custom]

\reftitle{References}

\end{document}